\documentclass[fleqn,usenatbib]{mnras}

\usepackage{newtxtext,newtxmath}

\usepackage[T1]{fontenc}

\DeclareRobustCommand{\VAN}[3]{#2}
\let\VANthebibliography\thebibliography
\def\thebibliography{\DeclareRobustCommand{\VAN}[3]{##3}\VANthebibliography}

\usepackage{graphicx}	
\usepackage{amsmath}	
\usepackage{amssymb}	
\usepackage{comment}	

\title[Proper motion in CW searches]{Effects of proper motion of neutron stars on continuous gravitational-wave searches}

\author[P. B. Covas]{
P. B. Covas,$^{1}$\thanks{E-mail: jb.covas@uib.es (PBC)}
\\
$^{1}$Universitat de les Illes Balears, IAC3---IEEC, E-07122 Palma de Mallorca, Spain
}

\date{Accepted XXX. Received YYY; in original form ZZZ}

\pubyear{2020}

\begin{document}
\label{firstpage}
\pagerange{\pageref{firstpage}--\pageref{lastpage}}
\maketitle

\begin{abstract}
All-sky and directed continuous gravitational-wave searches look for signals from unknown asymmetric rotating neutron stars. These searches do not take into account the proper motion of the neutron star, assuming that the loss of signal-to-noise ratio caused by this is negligible and that no biases in parameter estimation are introduced. In this paper we study the effect that proper motion has on continuous wave searches, and we show for what regions of parameter space (frequency, proper motion, sky position) and observation times this assumption may not be valid. We also calculate the relative uncertainty on the proper motion parameter estimation that these searches can achieve.
\end{abstract}




\section{Introduction}
\label{sec:intr}


Continuous waves (CWs) are long-lasting and almost monochromatic gravitational waves that can be emitted by rotating neutron stars if they are asymmetric around their rotation axis. 
These asymmetries can be supported either by elastic or magnetic deformations, as recently summarized in \citep{CWReview}. Many searches for CWs have been done in the past, looking both for CWs from known pulsars and from unseen neutron stars in our galaxy, both from known locations such as the galactic center or from all the sky \citep{O21,O2AllSky,O23}. These searches have not reported a CW detection, placing bounds on the maximum gravitational-wave amplitude.

The optimal frequentist technique to uncover a signal buried in Gaussian noise is the matched filtering, where the data obtained by ground-based detectors such as Advanced LIGO \citep{AdvancedLIGO} is correlated with a theoretical waveform. These waveforms are generated after a signal model has been assumed, and when this model does not accurately describe the true waveform the signal may not be found. 
The typical CW signal model takes into account the Doppler modulation produced by Earth's rotation and orbit around the solar system barycenter (SSB), and the spin-down of the neutron star produced by the emission of electromagnetic and gravitational radiation. In order to describe this model, 4 amplitude parameters (the amplitude $h_0$, initial phase $\phi_0$, polarization angles $\iota$ and $\psi$) and $3+s$ phase parameters (initial frequency $f_0$, sky position $\alpha$ and $\delta$, and $s$ spin-down parameters such as $f_1$ and $f_2$) are used. When the neutron star is in a binary system more parameters are needed in order to take into account the Doppler modulation produced by the motion around the binary barycenter, where for the general case 5 additional parameters are needed (3 for the circular orbit case). Searches for CWs from known pulsars only need to perform the matched filtering once, since all the phase parameters that describe the waveform (without taking into account the so-called amplitude parameters) 
are previously known. On the other side, searches for CWs from unknown neutron stars have to calculate the matched filter over many different waveforms, which correspond to different combinations of the unknown phase parameters describing the source. 

When the signal model does not completely describe the signal (such as when the spin-down of the source is neglected), 
two different effects will take place: 
\begin{enumerate}
    \item The mismatch (loss of signal-to-noise ratio) produced by using an incorrect signal model will lower the probabilities of detection. 
    \item Even if the signal is not missed, the estimated parameters will be somewhat biased, which may difficult further confirmation of the source such as from a complementary electromagnetic detection.
\end{enumerate}

There are several physical processes that when unaccounted for may 
render the usual CW signal model incomplete, 
such as spin-wandering \citep{WanderingSpin}, the presence of glitches \citep{GlitchesStat}, timing noise \citep{TimingNoise}, or proper motion. 
In this paper we aim to quantify 
the effects produced on CW searches when proper motion is neglected.

Neutron stars are known to be high velocity objects \citep{StatProp}, and their proper motion has been measured only for less than 400 pulsars. 
Searches from known pulsars take into account the proper motion information if available, while searches for unknown neutron stars do not search over the two parameters that characterize the proper motion, assuming that the mismatch produced by this is negligible. In this study we derive analytical expressions that are able to estimate the mismatch produced by this assumption, and we find that for high frequencies and integration times longer than a year this may cause a large loss of signal-to-noise ratio.  

Prior to this paper, only one attempt to quantify the ability to measure proper motion by CW searches was reported in \citep{KrolakPE}, where a single example was treated: an integration time of four months, with only a single value of the proper motion and frequency. In that study it was obtained that the median relative uncertainty of the proper motion estimators was around 40 per cent. In this paper we quantify the ability to measure proper motion with more examples and by using a Bayesian MCMC procedure, instead of the analytical Fisher matrix. 

This paper is structured as follows: in section \ref{sec:phase} we develop the phase model and its dependence on the proper motion of the neutron star; in section \ref{sec:propmot} we give a summary of the measured proper motions of known pulsars; in section \ref{sec:mismatch} 
we calculate the loss of signal-to-noise ratio and biases when the proper motion parameters are neglected, and we present analytical equations that can predict the mismatch; 
in section \ref{sec:followup} we 
show the relative uncertainty that the proper motion estimators can achieve; in section \ref{sec:conclusions} we present our conclusions.

\section{Phase model}
\label{sec:phase}

A standard CW signal is described by the following equation \citep{Fstat}:
\begin{align}
    h(t) = h_0[F_+(t,\psi, \hat{n})\frac{1+\cos{\iota}}{2} \cos{\phi(t)} + F_{\times}(t,\psi,\hat{n}) \cos{\iota} \sin{\phi(t)} ],
    \label{eq:h0t}
\end{align}
where $F_+$ and $F_{\times}$ are the antenna patterns of the detectors (which can be found in \citep{Fstat}) for the two different gravitational-wave polarizations, $t$ is the time at the detector frame, the inclination angle $\iota$ is the angle between the neutron star angular momentum and the observer's sky plane, $\psi$ is the wave polarisation angle, $\phi(t)$ is the phase of the signal and $h_0$ is the amplitude of the signal. 
The signal given by equation \eqref{eq:h0t} is described by 4 amplitude parameters ($h_0$, $\iota$, $\psi$, $\phi_0$) and 3 ($f_0$, $\alpha$, $\delta$) $+ s$ phase parameters, where s is the number of spin-down/up parameters.



The rotational phase of a neutron star is usually described with a Taylor approximation around a reference time, where the different orders of the approximation represent frequency derivatives that are present due to the emission of electromagnetic and gravitational waves. 
For most of the known pulsars, only one frequency derivative is needed to describe this phase. We assume that the gravitational-wave phase equals two times the rotational phase, thus being described by:
\begin{align}
    \phi(\tau) = \phi_0 + 2\pi \sum_{k=0}^{s} f_k \frac{(\tau-t_r)^{k+1}}{(k+1)!},
    \label{eq:phaseevosource}
\end{align}
where we define $f_k$ as the kth-order  gravitational-wave frequency given at reference time $t_r$, while $\phi_0$ is an initial phase. 

To relate the phase in the source frame $\phi(\tau)$ to the phase in the detector frame $\phi(t)$, 
a timing relation that takes into account relativistic effects is developed in \citep{Fstat}. There it is shown that after performing another Taylor approximation, the most important terms affecting the phase are:
\begin{align}
  \phi(t) \cong \phi_{0} + 2 \pi \sum_{k=0}^{s} f'_k \frac{(t-t_r)^{k+1}}{(k+1) !} + \frac{2 \pi}{c} \hat{n} \cdot \vec{r} \sum_{k=0}^{s-1} f'_k \frac{(t-t_r)^{k}}{k !}
    \label{eq:phaseevosource2}
\end{align}
where $\vec{r}$ is the position of the detector with respect to the SSB and $\hat{n}$ is the position of the source in sky, given by $\hat{n}(t)=[\cos{\alpha(t)}\cos{\delta(t)},\sin{\alpha(t)}\cos{\delta(t)},\sin{\delta(t)}]$ where the two sky coordinates are described by (in equatorial coordinates):
\begin{align}
    \alpha(t) &= \alpha_0 + \mu_{\alpha}(t-t_r) \\
    \delta(t) &= \delta_0 + \mu_{\delta}(t-t_r),
\end{align}
where $\alpha_0$ and $\delta_0$ are the sky positions at reference time $t_r$, and $\mu_{\alpha}$ and $\mu_{\delta}$ are the proper motions in the right ascension and declination. As explained in \citep{Fstat}, the frequencies $f_k'$ appearing in equation \eqref{eq:phaseevosource2} are not equal to the frequencies $f_k$ in the source frame, differing by a constant offset. 

The source vector $\hat{n}(t)$ can be approximated by a Taylor expansion  
around $t_r$ up to first order in time:
\begin{align}
\hat{n}(t) & \approx \hat{n}(t_r) + \hat{\dot{n}}(t_r)(t-t_r) \nonumber \\
 &= [\cos{\alpha_0}\cos{\delta_0},\sin{\alpha_0}\cos{\delta_0},\sin{\delta_0}] \nonumber \\ 
 & + (t-t_r)[-\mu_{\alpha}\sin{\alpha_0}\cos{\delta_0} - \mu_{\delta}\cos{\alpha_0}\sin{\delta_0}, \nonumber \\
 & \mu_{\alpha}\cos{\alpha_0}\cos{\delta_0} - \mu_{\delta}\sin{\alpha_0}\sin{\delta_0}, \mu_{\delta}\cos{\delta_0}] .
\end{align}
It can be seen that for values of proper motion smaller than $\sim 10^{-14}$ rad/s, higher-order corrections do not have an important contribution for integration times of the order of a few years. 

The detector position is given by the sum of an Earth barycenter component (assumed to be circular) and the barycenter-to-detector component: $\vec{r} (t) = \vec{r}_O(t) + \vec{r}_d(t)$, described by:
\begin{align}
\vec{r}_O (t) &= R_{ES}[\cos{\left(\phi_O + \Omega_O (t-t_r)\right)}, \nonumber \\ & \cos{\epsilon}\sin{\left(\phi_O + \Omega_O (t-t_r)\right)}, \sin{\epsilon}\sin{\left(\phi_O + \Omega_O (t-t_r)\right)}] \nonumber \\
\vec{r}_d (t) &= R_E[\cos{\lambda}\cos{(\phi_r + \Omega_r (t-t_r))}, \nonumber \\ &\cos{\lambda}\sin{(\phi_r + \Omega_r (t-t_r))},\sin{\lambda}],
\end{align}
where $\Omega_O$ is the orbital angular velocity, $\Omega_r$ is the rotational angular velocity, $R_{ES}$ is the mean distance between the SSB and Earth's barycenter, $R_E$ is the distance between Earth's barycenter and the detector, $\epsilon$ is the ecliptic angle, $\lambda$ is the latitude of the detector, and $\phi_O$ and $\phi_r$ are initial phases. For the following analytical calculations, we will assume that the vector $\vec{r}(t)$ only consists on the first term $\vec{r}_O(t)$, since $R_{ES} \gg R_E$ and the effects produced by the rotational term can be neglected, as shown in the Appendix (although all the codes used in this paper use the full $\vec{r}$).





\section{Proper motion of neutron stars}
\label{sec:propmot}

\begin{figure}
\includegraphics[width=1.0\columnwidth]{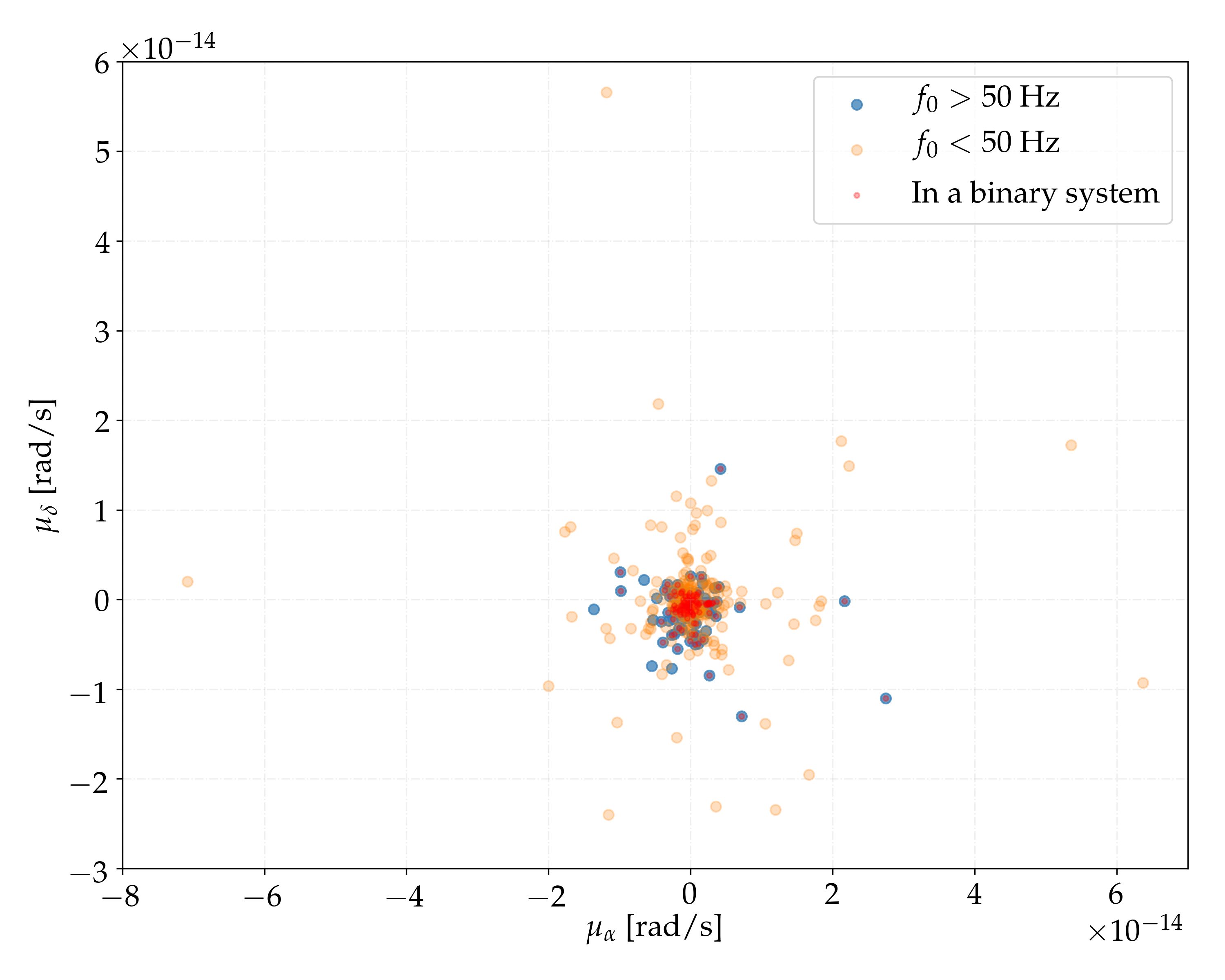}
\caption{Proper motion in right ascension and declination for 344 pulsars. The orange circles show 190 pulsars with gravitational-wave frequency less than 50 Hz, while blue circles show the other 154 pulsars. Pulsars in a binary system are shown with a red point. Data taken from \citep{ATNFF} and downloaded with \citep{psrqpy}.}
\label{fig:ProperMotions}
\end{figure}

Pulsars are known to have high spatial velocities, reaching up to 1500 km/s \citep{FastNS}. These proper motions are measured through electromagnetic detections of neutron stars, mainly by using three different mechanisms: pulsar timing \citep{TEMPO,PTA}; comparison between sky positions at different epochs \citep{CrabAccurateUncertainties}; scintillation \citep{Scintillation1,Scintillation2}. Proper motion has been measured only for 344 pulsars, shown in figure \ref{fig:ProperMotions}. It can be seen that $|\mu_{\alpha}|$ is above $10^{-14}$ for 30 pulsars and above $10^{-15}$ for 212 pulsars, while $|\mu_{\delta}|$ is above $10^{-14}$ for 18 pulsars and above $10^{-15}$ for 203 pulsars. This figure also shows that the majority (but not all) of pulsars with measured proper motion values with gravitational-wave frequencies higher than 50 Hz are located in a binary system.

The low number of pulsars with proper motions greater than $10^{-14}$ should be taken with care, since it is known that these measurements suffer from selection effects that bias them towards neutron stars with lower velocities, as discussed in \citep{BiasPropMotion}. Since the majority of the stellar progenitors of neutron stars belong to the Galactic plane, higher velocity pulsars spend less time within the detection volume of electromagnetic surveys than lower velocity ones. For this reason, even if the majority of the pulsars in the distribution seem to have values lower than $10^{-14}$, neutron stars with higher values cannot be dismissed. 


The highest velocities shown in figure \ref{fig:ProperMotions} are much higher than the usual space velocities of the progenitor stars of type O and B. Orbital velocities in binary systems only reach up to around 200 km/s, implying that an extra mechanism needs to be present. Several mechanisms have been proposed in order to explain these high velocities (although the dominant mechanism is not known): a post-natal electromagnetic rocket mechanism (due to asymmetric electromagnetic radiation when the magnetic dipole is displaced from the center of the star), which requires a high initial rotational frequency \citep{AsymmetricJet}; asymmetric radiation during the supernova, either of neutrinos \citep{AsymmetricNeutrinos}, hydrodynamical due to asymmetries in the mass ejection \citep{AsymmetricHydrodynamical}, or an asymmetric explosion of gamma-ray bursts \citep{AsymmetricGammaRay}. These high velocities could also be produced in dynamical interactions, for example with the supermassive black hole in the Galactic Center or within globular clusters. 

When detected, a proper motion measurement from a continuous wave signal could help to improve our understanding of the mechanism that gives rise to the observed proper motions, since gravitational waves may access a part of the galactic neutron star population that is hidden from the electromagnetic surveys. For example, a very high transverse velocity measurement can be used to constrain the physics of supernova core collapse. Furthermore, many proper motion measurements from CW detections could help to solve the hypothesis of whether there is a correlation between the proper motion vector and the angular momentum vector (spin-kick alignment) \citep{CorrelationPMAM}, since the latter can also be measured with CWs (represented by the $\psi$ and $\iota$ parameters). This correlation would be helpful in determining a specific mechanism for the production of such high proper motion values, since not all of the previous mentioned mechanisms predict the spin-kick alignment \citep{ProperMotionJet}. Moreover, measuring the proper motion of a neutron star allows the estimation of the birth site (after an age estimate is done), which can be used to associate the neutron star with a supernova remnant or a nebula.




\section{Mismatch}
\label{sec:mismatch}

As previously mentioned, all searches for CWs from unknown neutron stars (both all-sky and from known sky positions such as the galactic center) assume that the proper motion of the source is zero. In this section we calculate what is the expected loss of signal-to-noise ratio 
due to this assumption. 

In order to detect a signal, CW searches for unknown neutron stars calculate a detection statistic for all the different templates (combinations of phase parameters) that are searched. This detection statistic sorts the templates by the probability that a true astrophysical signal described by those parameters is present in the data. 
One of the most used detection statistics is the $\mathcal{F}$-statistic, which is the frequentist likelihood ratio maximized over the 4 amplitude parameters that define the CW signal, firstly developed in \citep{Fstat}. 
A signal is said to be detected (or saved for follow-up) if the $\mathcal{F}$-statistic value for some template is above a certain threshold, calculated from the false alarm probability that is created from the background noise.  
The expected value of the $\mathcal{F}$-statistic is related to the signal-to-noise ratio of the signal \citep{Fstat}:
\begin{align}
    \left< 2\mathcal{F} \right> = 4 + \rho^2(0),
    \label{eq:expectedF}
\end{align}
where $\rho^2(0)$ is the squared signal-to-noise ratio (SNR) when there is no mismatch, i.e. when the searched parameters are exactly equal to the astrophysical parameters.

Due to prohibitively high computational costs, all-sky and (some) directed searches calculate a semi-coherent detection statistic, where the data is separated in shorter segments and phase coherence is only demanded within each of these segments, but not between them. The expected value of the semi-coherent $\mathcal{F}$-statistic is:
\begin{align}
    \left< 2\mathcal{\tilde{F}} \right> = 4N + \tilde{\rho}^2(0),
    \label{eq:expectedF2}
\end{align}
where $\tilde{\rho}^2(0)$ is them sum of the signal-to-noise ratios of each segment. Since the values of the searched parameters will never be exactly equal to the parameters of the astrophysical signal, a fraction of the signal-to-noise ratio is not recovered. The mismatch $m$ describes the amount of squared SNR that is lost due to not searching at exactly the signal parameters, and it is given by:
\begin{align}
    m = \frac{\rho^2(0) - \rho^2(m)}{\rho^2(0)},
    \label{eq:mismatch}
\end{align}
ranging from 0 (fully recovered SNR) to 1 (no recovered SNR). The mismatch lowers the obtained $\mathcal{F}$-statistic value, implying that a signal that would be detectable without mismatch may not be recovered.

The mismatch can be estimated by doing a Taylor expansion of the likelihood ratio around the signal parameters, where it attains a maximum. Usually, only the second-order term is kept: 
\begin{align}
    m \approx g_{ij} (\Theta) d\Theta^i d\Theta^j + \mathcal{O}(d\Theta^3),
    \label{eq:phasemetric}
\end{align}
where $g_{ij}$ is the parameter space metric ($i$ and $j$ run over the dimensions, given by the number of parameters) and $\Theta$ represents the different parameters, such as frequency or sky positions. This approximated mismatch is unbounded and can be higher than 1, and from previous studies it is known that this approximation highly overestimates the actual mismatch for mismatches higher than $\sim 0.3$ \citep{PrixMultiMetric}, a fact that was also studied in \citep{ExtendedMetric}, which further analyzed where the metric approximation breaks down. 


\subsection{Parameter bias and expected mismatch}
\label{subsec:bias}

Since all-sky and directed searches do not search over the proper motion parameters,  
some mismatch will always be present (even when an infinitely fine grid over the other searched parameters is used), but usually it is assumed that this mismatch is much lower than the mismatch produced by the other parameters. 
A similar situation was discussed in \citep{GlitchesStat}, where the mismatch produced by the presence of glitches in the signal was studied. 
As noted there, the template that attains the minimum mismatch will not be located at the true signal parameters, since there will be a shifted template combination that minimizes the effect of the missing proper motion parameters. The minimum mismatch and displaced parameters can be estimated by minimizing the mismatch function given by equation \eqref{eq:phasemetric} (where we have separated the parameters between searched parameters $\lambda$ and non-searched proper motion parameters $\Lambda$):
\begin{align}
    m = g_{ij} \Delta \lambda^i \Delta \lambda^j + g_{kl} \Delta \Lambda^k \Delta \Lambda^l + 2 g_{ik} \Delta \lambda^i \Delta \Lambda^k,
    \label{eq:mismatch0}
\end{align}
where the indices $k$ and $l$ only go from 0 to 1 (the two proper motion parameters). The minimum mismatch will be obtained at non-zero displacements of the searched $\lambda$ parameters. These displacements can be estimated by minimizing the previous mismatch equation:
\begin{align}
    \frac{ \partial m}{\partial \Delta \lambda^i} &= 0 \longrightarrow \Delta_{min} \lambda^j = - g_{ij}^{-1} g_{ik} \Delta \Lambda^k \\
    m_{min} &= g_{ij} g^{-1}_{ij} g^{-1}_{ji} g_{ik}g_{jk} (\Delta \Lambda^k)^2 + g_{kl} \Delta \Lambda^k \Delta \Lambda^l \nonumber \\ &- 2 g_{ik} g^{-1}_{ji} g_{jk} (\Delta \Lambda^k)^2  \nonumber \\
    &= g_{kl} \Delta \Lambda^k \Delta \Lambda^l - g^{-1}_{ij} g_{ik} g_{jk} (\Delta \Lambda^k)^2.
    \label{eq:minmismatch}
\end{align}
As mentioned in \citep{GlitchesStat}, this expression is only valid for displacements that would produce a mismatch lower than $\sim 0.3$, since for higher mismatches the second-order Taylor approximation is not valid.

For example, if the unknown searched parameter was $f_0$, 
and $\mu_{\alpha}$ was the unknown non-searched parameter, these expressions would be:
\begin{align}
    m &= g_{f_0f_0}(\Delta f_0)^2 + g_{\mu_{\alpha} \mu_{\alpha}}(\Delta \mu_{\alpha})^2 + 2 g_{f_0 \mu_{\alpha}} \Delta f_0 \Delta \mu_{\alpha} \nonumber \\
    \Delta_{min} f_0 &= - \frac{g_{f_0 \mu_{\alpha}}}{g_{f_0 f_0}} \Delta \mu_{\alpha} \nonumber \\
    m_{min} &= g_{f_0f_0} \left(\frac{g_{f_0 \mu_{\alpha}}}{g_{f_0 f_0}}\right)^2 (\Delta \mu_{\alpha})^2 + g_{\mu_{\alpha} \mu_{\alpha}}(\Delta \mu_{\alpha})^2 \nonumber \\ & \, - 2 g_{f_0 \mu_{\alpha}} \frac{g_{f_0 \mu_{\alpha}}}{g_{f_0 f_0}} (\Delta \mu_{\alpha})^2 \nonumber \\    & = \left( g_{\mu_{\alpha} \mu_{\alpha}} - \frac{g^2_{f_0 \mu_{\alpha}}}{g_{f_0 f_0}} \right) (\Delta \mu_{\alpha})^2.
\end{align}
It can be seen that the mismatch is reduced due to the second negative term of the last equation, as compared to the simple case where $\Delta f_0 = 0$: 
\begin{align}
  m_N = g_{\mu_{\alpha} \mu_{\alpha}}(\Delta \mu_{\alpha})^2.
\end{align}
We remark that these are the minimum mismatches that would be obtained if we searched over an infinitely finely spaced template bank over the frequency, spin-down, and sky positions. In a more realistic scenario, the mismatch will always be bigger than this minimum mismatch.


In order to calculate these parameter displacements and minimum mismatches, we use a modified (which includes the two proper motion components) version of the \textit{UniversalDopplerMetric} code from the LALSuite repository \citep{lalsuite}, which is able to calculate the metric components by computing equation 87 from \citep{PrixMultiMetric}. After finding all the metric components we can calculate the parameter displacements and the minimum mismatch.

\begin{figure}
\includegraphics[width=1.0\columnwidth]{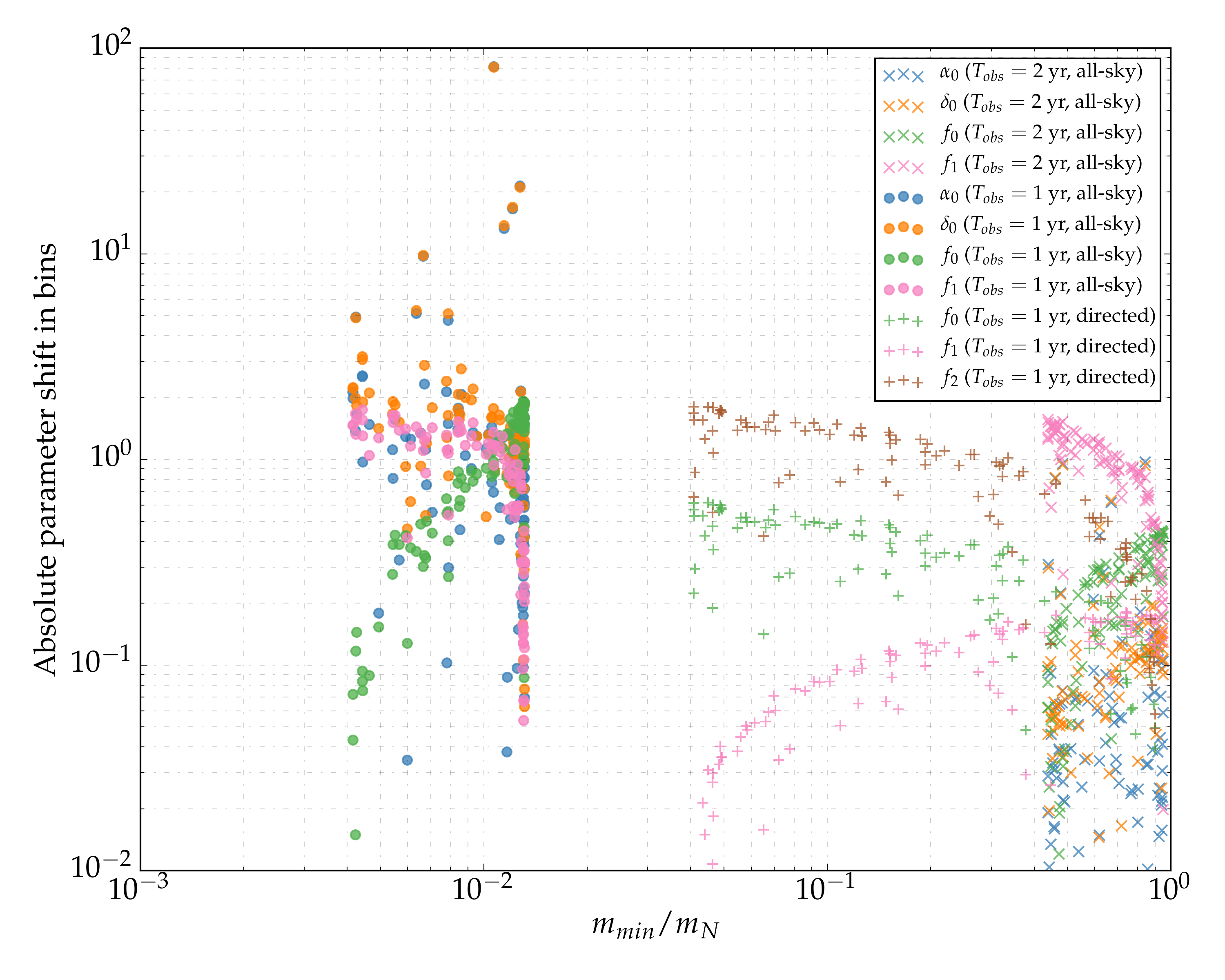}
\caption{Absolute shift of the different parameters compared to the ratio of minimum mismatch given by equation \eqref{eq:minmismatch} and simple mismatch $m_N$ given by equation \eqref{eq:mismatch0} when $\Delta \lambda^i = 0$ for all $i$. The bins have been defined as $\sqrt{0.1/g_{\lambda_i \lambda_i}}$. These results are for a fully coherent search, where the reference time has been defined as $t_{mid}$.}
\label{fig:Bias}
\end{figure}

The parameter biases are shown in figure \ref{fig:Bias}, together with the fraction of minimum mismatch compared to the $m_N$ mismatch. We have simulated signals from neutron stars with isotropic sky positions and orientations, with frequencies from 100 to 1500 Hz. It can be seen that the fraction between the minimum and $m_N$ mismatches highly depends on the total observing time $T_{obs}$. When there is more reduction in the mismatch, the recovered parameters deviate more from the true parameters: the sky positions can differ from the true sky positions by more than 5 bins. For the 1 year case, the sky positions are the parameters that are more biased, while for the 2 years search the first frequency derivative is more biased. For the directed search, the second frequency derivative has the highest bias.  
These results have been obtained by setting the reference time $t_r$ to the middle of the observation time. The minimum mismatch $m_{min}$ is an invariant quantity with respect to $t_r$, but the sizes of the parameter bias are greatly incremented when using other reference times, such as the initial or ending times of the observation. 
We remark the fact that the biases shown in this figure for $f_0$, $f_1$, and $f_2$ are between the recovered value and the modified primed frequencies that appear in equation \eqref{eq:phaseevosource2}, not between the recovered values and the source-frame frequencies.

The figure also shows than the reduction of minimum mismatch is smaller for directed searches, since the sky position is fixed (although a second spin-down parameter is also searched). This figure clearly shows that 
biases created by assuming the proper motion to be zero can be much larger than the typical resolution of the search. 
In order to confirm these calculations, we have compared the obtained results with the mismatch obtained when calculating the $\mathcal{F}$-statistic values at both the signal and the displaced parameters, using the \textit{lalapps\_ComputeFstatistic\_v2} code (also part of the LALSuite repository). This procedure has returned the same mismatch results as obtained with the \textit{UniversalDopplerMetric} code.


\begin{figure*}
\centering
\includegraphics[width=1.0\columnwidth]{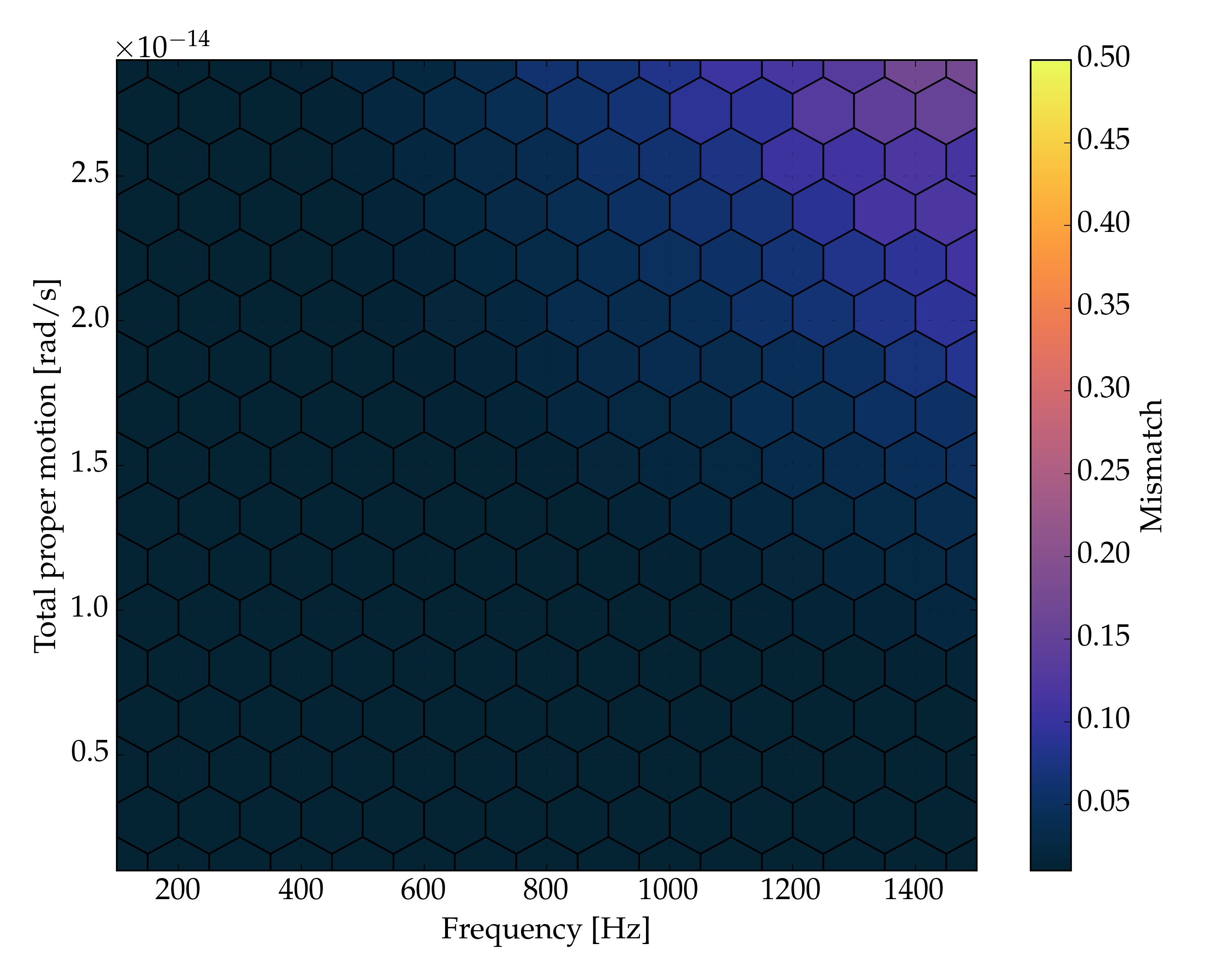}
\includegraphics[width=1.0\columnwidth]{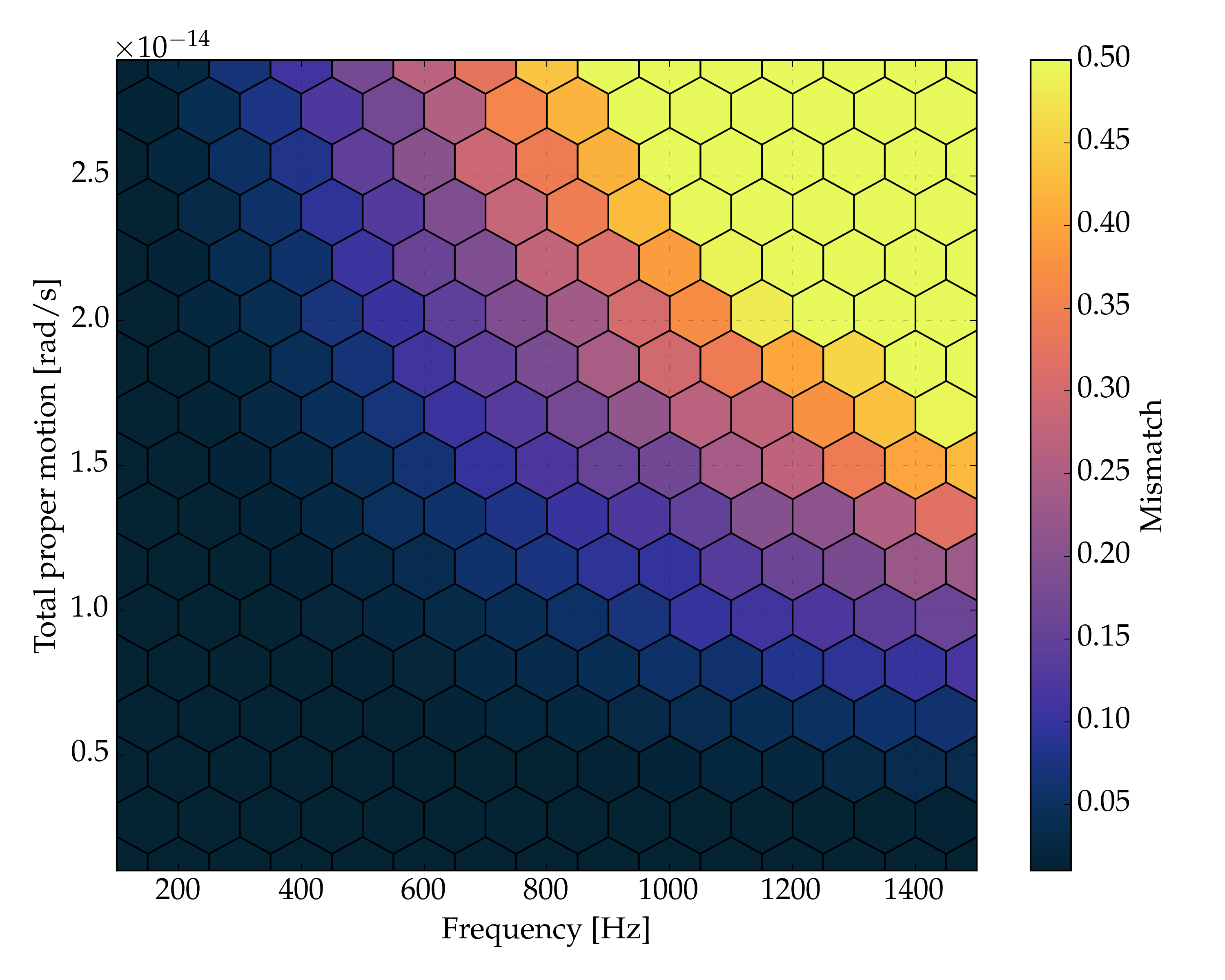}
\caption{These plots show the average (between 100 signals for each cell) minimum mismatch given by equation \eqref{eq:minmismatch} as a function of frequency and total proper motion for two different coherent times: 1.5 years (left), and 2 years (right). 
The reference time for these searches has been selected as the middle of the observing time.
}
\label{fig:Mism3D}
\end{figure*}

We have repeated these calculations by varying the frequency and total proper motion components, in order to study at which regions of parameter space will the minimum mismatch exceed a certain threshold value. The results are shown in figure \ref{fig:Mism3D}, where two different plots are shown for two different integration times. Each cell of the plot is made by averaging the results from 100 signals distributed with an isotropic sky position and random amplitude parameters (producing SNRs between 10 and 1000). It can be seen that the mismatch increases with the frequency and with the total proper motion value, and also with the coherent integration time. The maximum proper motion value in these plots is $2.9 \times 10^{-14}$ rad/s, although as discussed previously unknown neutron stars could attain even higher proper motion values. For observing times smaller than 1 year, no minimum mismatches above 0.01 have been obtained. From these plots it can be seen that when doing a search with a coherent time longer than a year and not searching the proper motion parameters, there is a non-negligible probability of having a high mismatch and missing a signal for gravitational-wave frequencies greater than $\sim 600$ Hz. At lower frequencies, if the total proper motion is higher than $3 \times 10
^{-14}$ rad/s, it can be inferred that for coherent integration times longer than 2 years the mismatch could also be non-negligible. 

\begin{figure}
\includegraphics[width=1.0\columnwidth]{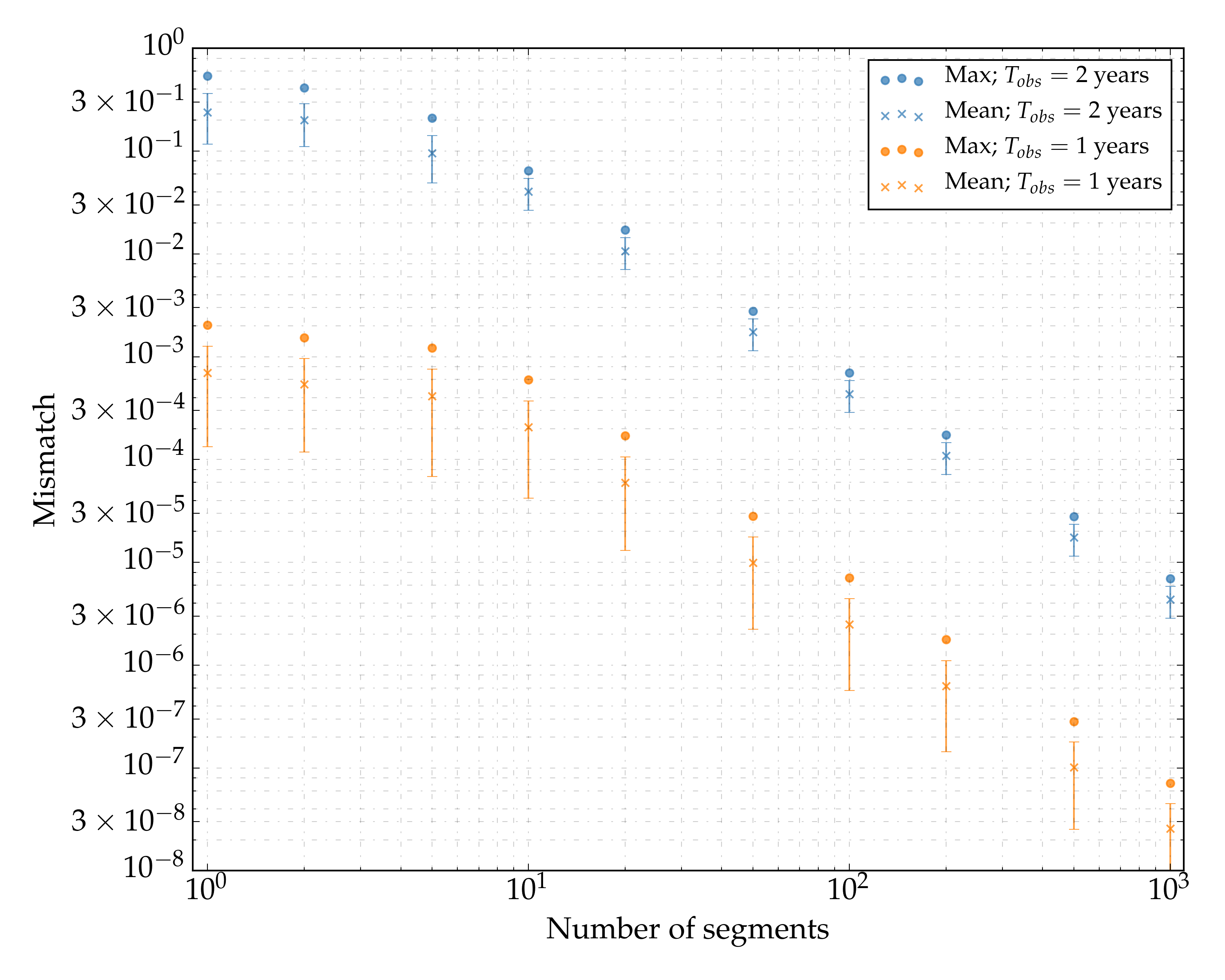}
\caption{Scaling of the minimum mismatch with respect to the number of segments in a semi-coherent search. Circles show the maximum mismatch, while crosses show the average result from 100 signals belonging to one cell of figure \ref{fig:Mism3D} (with 1-$\sigma$ error bars). Blue points show the results for an observation time of 2 years, while orange points show the results for 1 year.}
\label{fig:semicoh}
\end{figure}

The mismatches for the semi-coherent case are shown in figure \ref{fig:semicoh}. These results belong to one single cell of figure \ref{fig:Mism3D}, but very similar results with the same scaling are obtained for all other cells. The semi-coherent metric components are obtained by averaging the different coherent integrations, where for each them the starting time $t_i$ will be different. The figure shows that for less than 5 segments the mismatch is comparable to the fully coherent case, but when there are more segments the mismatch quickly reduces to negligible amounts. 

The previous calculations show that we expect high mismatches only for observation times longer than a year. Although the past O1 and O2 observing runs have been of approximately 4 and 9 months respectively, the newest O3 run has lasted for about a year, and future observing runs are planned to be longer than a year, as discussed in \citep{ObsScen}. Furthermore, sometimes data from different observing runs has been combined in order to follow-up candidates from a search, as shown in \citep{FollowupEinstein}. These considerations show that for future analysis an explicit search over the proper motion parameters might be needed in order to safely lower the probabilities of missing a signal. Since high mismatches are obtained only for small number of segments, we expect that an explicit search over the proper motion parameters might only be needed at the last stages of a typical CW follow-up procedure, where the initial stages have a large number of segments and subsequent stages reduce the number of segments. The number of candidates is reduced at each stage of the follow-up, and at the last stages a very small amount of candidates remains above the threshold. For this reason, the increase in computational cost produced by searching over two extra parameters would not highly increase the final cost of a follow-up procedure, since it would only be needed at the last stages. 

A caveat of these results is that they have been obtained for 
duty cycles of 1 (i.e. simulated Gaussian data without gaps), while realistic data from gravitational-wave detectors always has duty cycles smaller than 1. We leave for future work an estimation of the effect that this would have on our results, but we believe that the calculated mismatches would not be reduced by more than a $\sim 0.75$ factor for realistic duty cycles.



\subsection{Derivation of the proper motion coherent metric components}




In order to quickly estimate the mismatch that will be present when the proper motion is neglected, an analytical equation is needed. From the previous subsection it can be seen that we need to calculate all the components of the parameter space metric that are related to the proper motion, such as $g_{\mu_{\alpha} \mu_{\alpha}}$. These metric components can be used to estimate the mismatch or to construct a bank of templates with a desired resolution.

The phase metric approximation is used to obtain these components, where the amplitude parameters are taken as constant and only the phase parameters are taken into account \citep{PrixMultiMetric}. Within this approximation, the metric components are given by:
\begin{align}
    g_{ij} = 
    \langle \partial_i \phi (\Theta) \partial_j \phi (\Theta) \rangle - \langle \partial_i \phi (\Theta) \rangle \langle \partial_j \phi (\Theta) \rangle,
\label{eq:metricdef}
\end{align}
where 
\begin{align}
\partial_i \phi (\Theta) = \frac{ \partial \phi}{\partial \Theta_i} \bigg\rvert_{\Theta_i=\Theta_s} \quad \text{   and   } \quad \langle \phi \rangle = \frac{1}{T} \int_{t_i}^{t_i+T} \phi(t) dt,
\end{align}
where $t_i$ is the starting time of the integral.


The $g_{\mu_{\alpha} \mu_{\alpha}}$ metric component is explicitly derived in the Appendix. Here we show all the metric components related to the proper motion, where we only keep the terms that depend on the highest order of the coherent integration time $T$ (when the reference time is defined as $t_{mid}$): 
\begin{align}
    g_{\mu_{\alpha}\mu_{\alpha}} &\approx \frac{4\pi^2 R_{ES}^2 f_0^2 T^2 }{24 c^2} [
    \cos^2{\delta_0}\sin^2{\alpha_0} + \cos^2{\epsilon}\cos^2{\alpha_0}\cos^2{\delta_0}], \nonumber \\
    g_{\mu_{\delta}\mu_{\delta}} &\approx \frac{4\pi^2 R_{ES}^2 f_0^2 T^2}{24c^2} [
    \cos^2{\alpha_0}\sin^2{\delta_0} + \cos^2{\epsilon}\sin^2{\delta_0}\sin^2{\alpha_0} \nonumber \\ & + \sin^2{\epsilon}\cos^2{\delta_0} - 2\cos{\epsilon}\sin{\epsilon}\cos{\delta_0}\sin{\delta_0}\sin{\alpha_0}], \nonumber \\
    g_{\mu_{\alpha}\mu_{\delta}} &\approx \frac{4\pi^2 R_{ES}^2 f_0^2 T^2}{24c^2} [(1-\cos^2{\epsilon})\cos{\alpha_0}\sin{\delta_0}\sin{\alpha_0}\cos{\delta_0} \nonumber \\ & + \cos{\epsilon}\sin{\epsilon}\cos{\alpha_0}\cos^2{\delta_0}], \nonumber \\
    g_{\mu_{\alpha}\alpha_0} &\approx 
    \frac{4\pi^2R_{ES}^2f_0^2}{2 \Omega_O c^2} \sin{\phi_O} \cos{\phi_O} \cos{T\Omega_O} [\cos^2{\delta_0}\sin^2{\alpha_0} \nonumber \\ - &\cos^2{\epsilon}\cos^2{\alpha_0}\cos^2{\delta_0} - \cos{\alpha_0}\sin{\alpha_0}\cos^2{\delta_0}\cos{\epsilon} (1 \nonumber \\ - &2\sin^2{\frac{T\Omega_O}{2}} \frac{\sin{\phi_O}}{\cos{\phi_O}} - 2\cos^2{\frac{T\Omega_O}{2}} \frac{\cos{\phi_O}}{\sin{\phi_O}})], \nonumber \\
    g_{\mu_{\delta}\alpha_0} &\approx 
    \frac{4\pi^2R_{ES}^2f_0^2}{2 \Omega_O c^2} \sin{\phi_O} \cos{\phi_O} \cos{T\Omega_O} [\cos{\delta_0}\sin{\delta}\cos{\alpha}\sin{\alpha_0} \nonumber \\ + &\cos^2{\epsilon}\sin{\alpha_0}\cos{\alpha_0}\cos{\delta_0}\sin{\delta_0} - \cos{\epsilon}\sin{\epsilon}\cos{\alpha_0}\cos^2{\delta_0} \nonumber \\ + &(\sin^2{\alpha_0}\cos{\delta_0}\sin{\delta_0}\cos{\epsilon} - \sin{\alpha_0}\cos^2{\delta_0}\sin{\epsilon} \nonumber \\ - &\cos^2{\alpha}\cos{\delta_0}\sin{\delta_0}\cos{\epsilon}) \nonumber \\ &(\frac{1}{2}  - \sin^2{\frac{T\Omega_O}{2}} \frac{\sin{\phi_O}}{\cos{\phi_O}} - \cos^2{\frac{T\Omega_O}{2}} \frac{\cos{\phi_O}}{\sin{\phi_O}})], \nonumber \\
    g_{\mu_{\alpha}\delta_0} &= g_{\mu_{\delta}\alpha_0}, \nonumber
\end{align}
\begin{align}
    g_{\mu_{\delta}\delta_0} &\approx 
    \frac{4\pi^2R_{ES}^2f_0^2}{2 \Omega_O c^2} \sin{\phi_O} \cos{\phi_O} \cos{T\Omega_O} [\cos^2{\alpha_0}\sin^2{\delta_0} \nonumber \\ - &\cos^2{\epsilon}\sin^2{\alpha_0}\sin^2{\delta_0} - \sin^2{\epsilon}\cos^2{\delta_0} \nonumber \\
    + &(\cos{\alpha_0}\sin{\alpha_0}\sin^2{\delta_0}\cos{\epsilon} - \cos{\alpha_0}\cos{\delta_0}\sin{\delta_0}\sin{\epsilon}) \nonumber \\ &(1  - 2\sin^2{\frac{T\Omega_O}{2}} \frac{\sin{\phi_O}}{\cos{\phi_O}} - 2\cos^2{\frac{T\Omega_O}{2}} \frac{\cos{\phi_O}}{\sin{\phi_O}})], \nonumber \\
    g_{\mu_{\alpha}f_0} &\approx
    \frac{4\pi^2 R_{ES}f_0T}{2 c\Omega_O} \sin{\frac{\Omega_O T}{2}} [-\cos{\delta_0}\sin{\alpha_0} \nonumber \\ & + \cos{\epsilon}\cos{\delta_0}\cos{\alpha_0}], \nonumber \\
    g_{\mu_{\alpha}f_1} &\approx 
    \frac{4\pi^2 R_{ES}f_0T^2}{4 c\Omega_O} \cos{\frac{\Omega_O T}{2}} [-\cos{\delta_0}\sin{\alpha_0}(1+\frac{1}{3}\sin{\phi_O}) \nonumber \\ & - \cos{\epsilon}\cos{\delta_0}\cos{\alpha_0} (1+\frac{1}{3}\cos{\phi_O})], \nonumber
\end{align}
\begin{align}
    g_{\mu_{\delta}f_0} &\approx 
    \frac{4\pi^2 R_{ES}f_0T}{2 c\Omega_O} \sin{\frac{\Omega_O T}{2}} [-\cos{\alpha_0}\sin{\delta_0} \nonumber \\ & - \cos{\epsilon}\sin{\delta_0}\sin{\alpha_0} + \sin{\epsilon} \cos{\delta_0} ], \nonumber \\
    g_{\mu_{\delta}f_1} &\approx 
    \frac{4\pi^2 R_{ES}f_0T^2}{4 c\Omega_O} \cos{\frac{\Omega_O T}{2}} [-\cos{\alpha_0}\sin{\delta_0}(1+\frac{1}{3}\sin{\phi_O}) \nonumber \\ & + (1+\frac{1}{3}\cos{\phi_O})(\cos{\epsilon}\sin{\delta_0}\sin{\alpha_0} - \sin{\epsilon}\cos{\delta_0}) ].
\end{align}
It can be noticed that these metric components depend on the sky position of the source and on its frequency, in a very similar way to the sky position metric components. 

For a search that has to cover all the sky, this would produce difficulties in the template bank construction (as explained in \citep{LatticeAllSky}), but, as argued before, the proper motion components only produce a noticeable mismatch for observation times longer than a year and with less than 5 segments. All-sky and directed searches only allow such long coherent times at the last stages of the follow-up procedure. In these stages the sky position of the source has already been determined with enough accuracy that it can be used as a constant input to the proper motion metric components. 

\begin{figure}
\includegraphics[width=1.0\columnwidth]{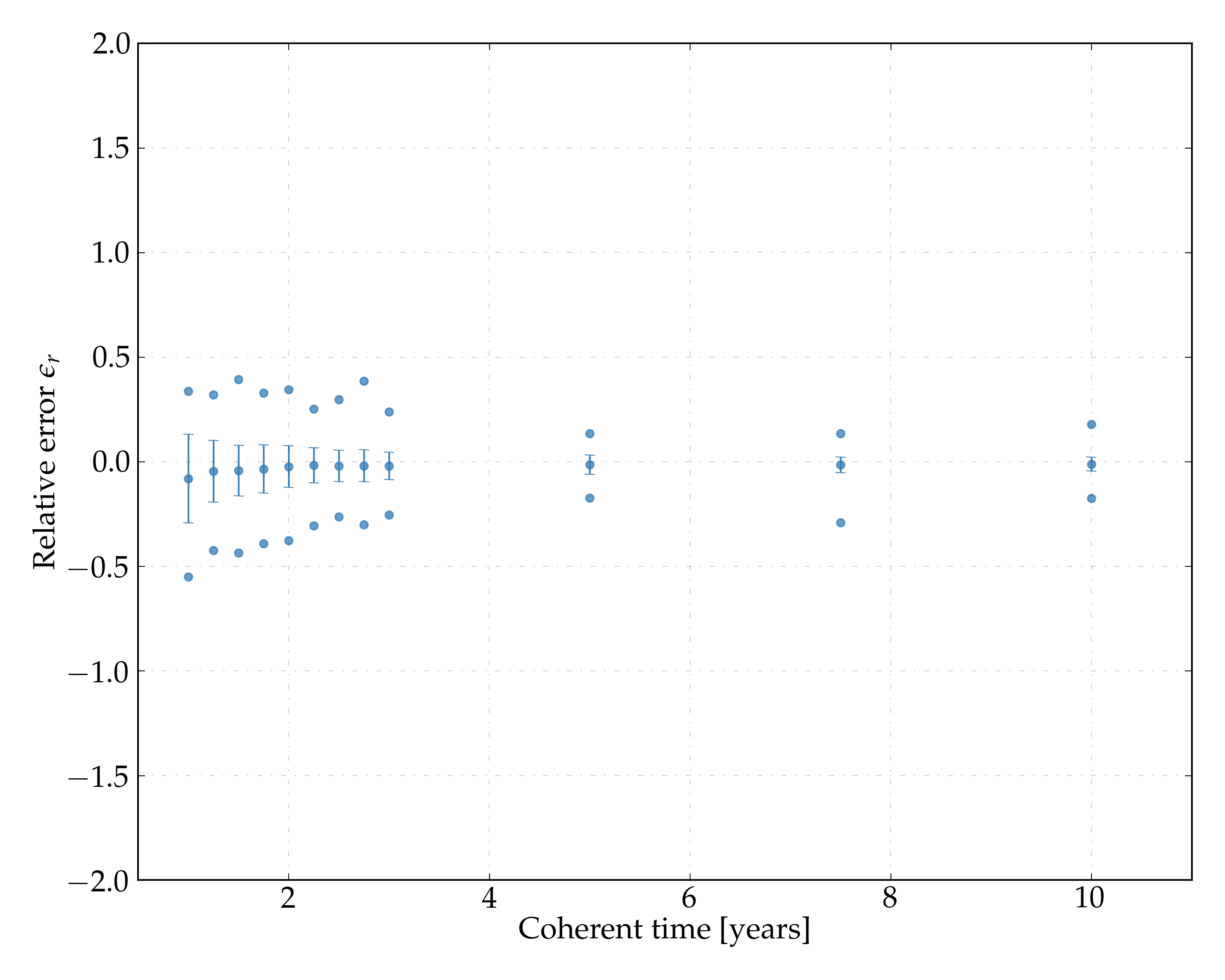}
\caption{Relative error given by equation \eqref{eq:relerror} as a function of the coherent integration time. For each time, the maximum, mean, and minimum points are shown, along with an error bar comprising one standard deviation.}
\label{fig:err}
\end{figure}

In order to validate the previous metric components, we calculate the relative error $\epsilon_r$ between the true and predicted mismatches, in order to study the behaviour of the mismatch:
\begin{align}
    \epsilon_r = 2\frac{m_0 - m}{m_0 + m},
    \label{eq:relerror}
\end{align}
where $m_0$ is the true mismatch obtained by calculating the $\mathcal{F}$-statistic values and $m$ is the mismatch predicted by the phase metric components. The relative error is negative when the predicted mismatch is higher than the true mismatch, meaning that we are overestimating the mismatch. 
The relative error is shown in figure \ref{fig:err} for a different number of coherent integration times ($10^6$ signals have been used for each coherent time). Overall, the errors that we obtain have similar values to the errors obtained in other papers where the metric components are estimated, such as \citep{LatticeAllSky}. It can be seen that the relative error decreases as the coherent time is increased, due to the approximated metric components being more accurate as the neglected terms are less important.


\subsection{Mismatch for frequency searches}

Many semi-coherent CW searches (such as the two Hough algorithms used in \citep{O2AllSky}) just track the frequency-time pattern of the signal, instead of searching the full signal given by equation \eqref{eq:h0t}. For these searches we can estimate if proper motion will produce non-negligible mismatch by calculating the difference between the true and searched tracks, a method which has been previously used to estimate the mismatch produced by higher-order spin-down terms or by neglected eccentricity in binary systems \citep{SkyHough,BSH}. 

\begin{figure}
\includegraphics[width=1.0\columnwidth]{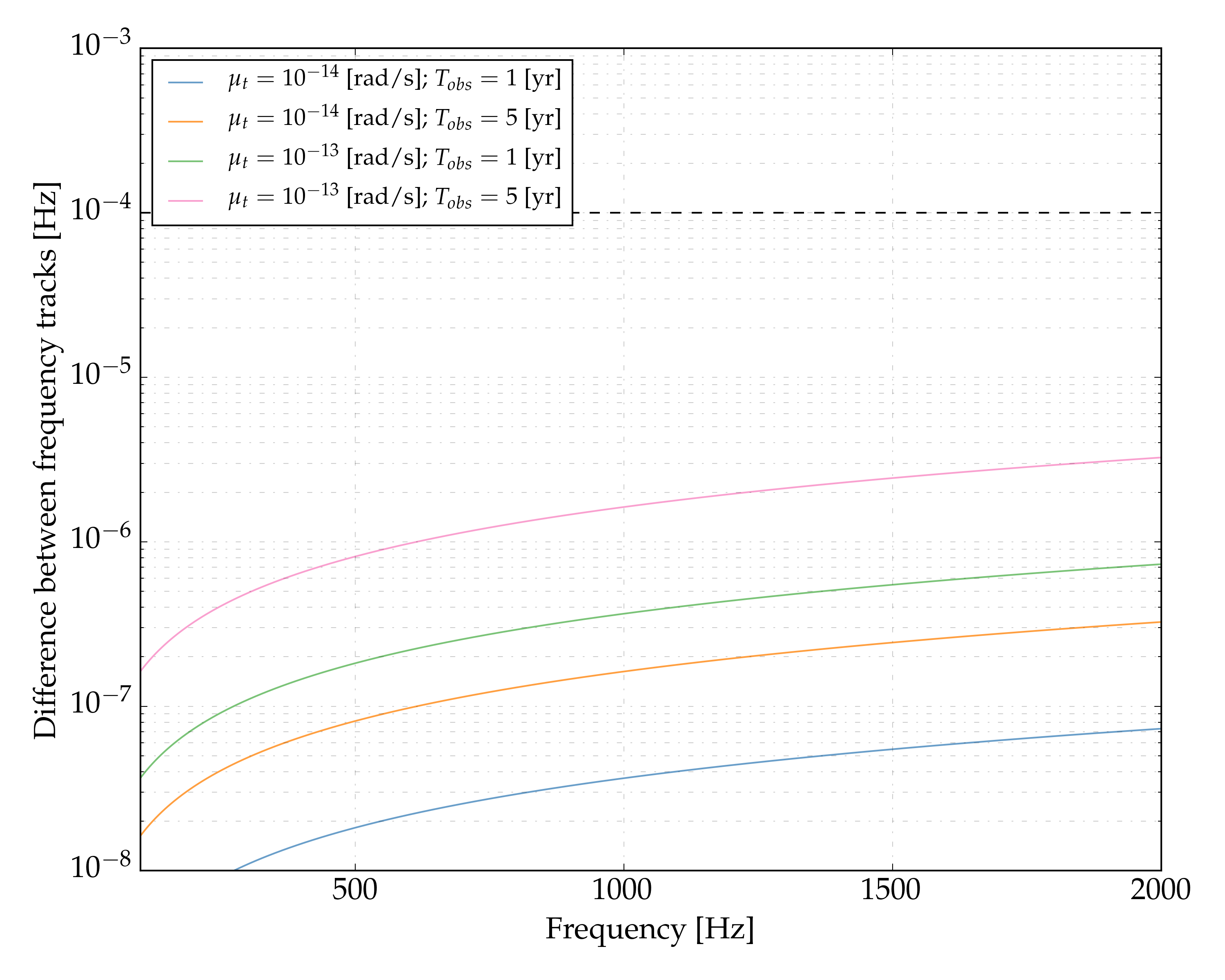}
\caption{Maximum difference between the frequency-time patterns given by equation \eqref{eq:SemiCoherentDiff}. Two different observing times and total proper motion values are shown.}
\label{fig:FrequencySearches}
\end{figure}

The frequency-time pattern (assuming $s=1$) is given by:
\begin{align}
    f_P(t) &= \frac{1}{2\pi} \frac{d\phi(t)}{dt} \nonumber \\ 
    &= f'_0 + f'_0\frac{\vec{v}(t)\cdot\hat{n}(t) + \vec{r}(t)\cdot\hat{v}_s(t)}{c} + f'_1 t \nonumber \\
    &\approx f'_0 + f'_0\frac{\vec{v}(t)\cdot \hat{n}_0}{c}  \nonumber \\ &+ f'_0\left( \frac{\vec{v}(t)\cdot \hat{\dot{n}}(t_r) (t-t_r)}{c} + \frac{\vec{r}(t)\cdot \hat{\dot{n}}(t_r)}{c} \right) + f'_1 t,
\end{align}
where $\hat{n}_0 \equiv \hat{n}(t_r)$, while when 
the proper motion parameters are not searched it is given by:
\begin{align}
  f(t) = f'_0 + f'_0\frac{\vec{v}(t)\cdot\hat{n}_0}{c} + f'_1 t.
\end{align}
The difference between the two frequency-time patterns is:
\begin{align}
  |f_P(t)- f(t)| &=  
  f'_0 \left( \frac{\vec{v}(t)\cdot \hat{\dot{n}}(t_r) (t-t_r)}{c} + \frac{\vec{r}(t)\cdot \hat{\dot{n}}(t_r)}{c} \right).
  \label{eq:SemiCoherentDiff}
\end{align}
A quick estimate shows that for signals with $f_0=1000$ Hz and an observation time of 1 year, the change in frequency will be smaller than $10^{-7}$ Hz for a total proper motion of $10^{-14}$ rad/s. The coherent time of these searches is usually less than 7200 s, which implies a frequency resolution of $df_0 \sim 10^{-4}$ Hz. This is shown in figure \ref{fig:FrequencySearches}, where four different traces are shown, for two different observation times and total proper motion values.

This shows that most semi-coherent searches are not able to detect the changes produced by proper motion since the calculated frequency evolution does not deviate by more than a frequency bin. This estimate is in agreement with the results shown in figure \ref{fig:semicoh}, where it can be seen that searches with a large number of segments (as is the case for these frequency tracking methods with short coherent times) do not have a mismatch higher than $10^{-3}$. 

\section{Proper motion parameter estimation}
\label{sec:followup}



In the previous section we have shown the mismatch present when the proper motion of the neutron star is assumed to be zero. In this section we study the accuracy that can be achieved when these parameters are searched.

A lower bound on the best achievable accuracy on the estimation of parameters can be obtained with the Fisher information matrix. As discussed in \citep{PrixMultiMetric}, the Fisher information matrix is related to the mismatch metric:
\begin{align}
    \Gamma_{ij} = \rho^2 g_{ij}.
\end{align}
The lower bound on the standard deviation is:
\begin{align}
    \sigma_i \geq \sqrt{\Gamma^{-1}_{ii}} = \frac{1}{\rho } \sqrt{g^{-1}_{ii}}.
    \label{eq:FisherMetric}
\end{align}

From this expression it is clear that the accuracy depends on the inverse of the SNR and the inverse of the metric. For this reason, the accuracy will depend on the frequency, the coherent time, and the SNR of the signal,  
and it is independent of the absolute value of the proper motion, i.e. we do not get better resolution for higher proper motions. 
On the other hand, the dependence of the accuracy on the number of detectors is present through the SNR of the signal: although the metric components are the same when more detectors are added (if the noise floors are the same), the SNR of the signal increments and so does the accuracy. 
From the expression it is also clear that a better accuracy will be achieved when only the proper motion parameters are unknown, since a search over other parameters will decrease the accuracy due to covariance. This means that in general the estimation of the proper motion parameters will be better for directed searches as compared to all-sky searches.

In order to empirically study the accuracy, we have modified the MCMC routine part of the \textit{pyfstat} repository \citep{Followup1,Followup2}. This software allows to do a $\mathcal{F}$-statistic search using a parallel tempered MCMC follow-up \citep{ptemcee1,ptemcee2}. We add the two proper motion parameters, and we do a coherent search with different coherent times (without data gaps) by searching an interval around the true parameters. We simulate an all-sky search where 4+2 parameters are searched, 
with two detectors of stationary Gaussian noise, and we add signals with a range of SNRs and with different Doppler parameters. 



\begin{figure}
\includegraphics[width=1.0\columnwidth]{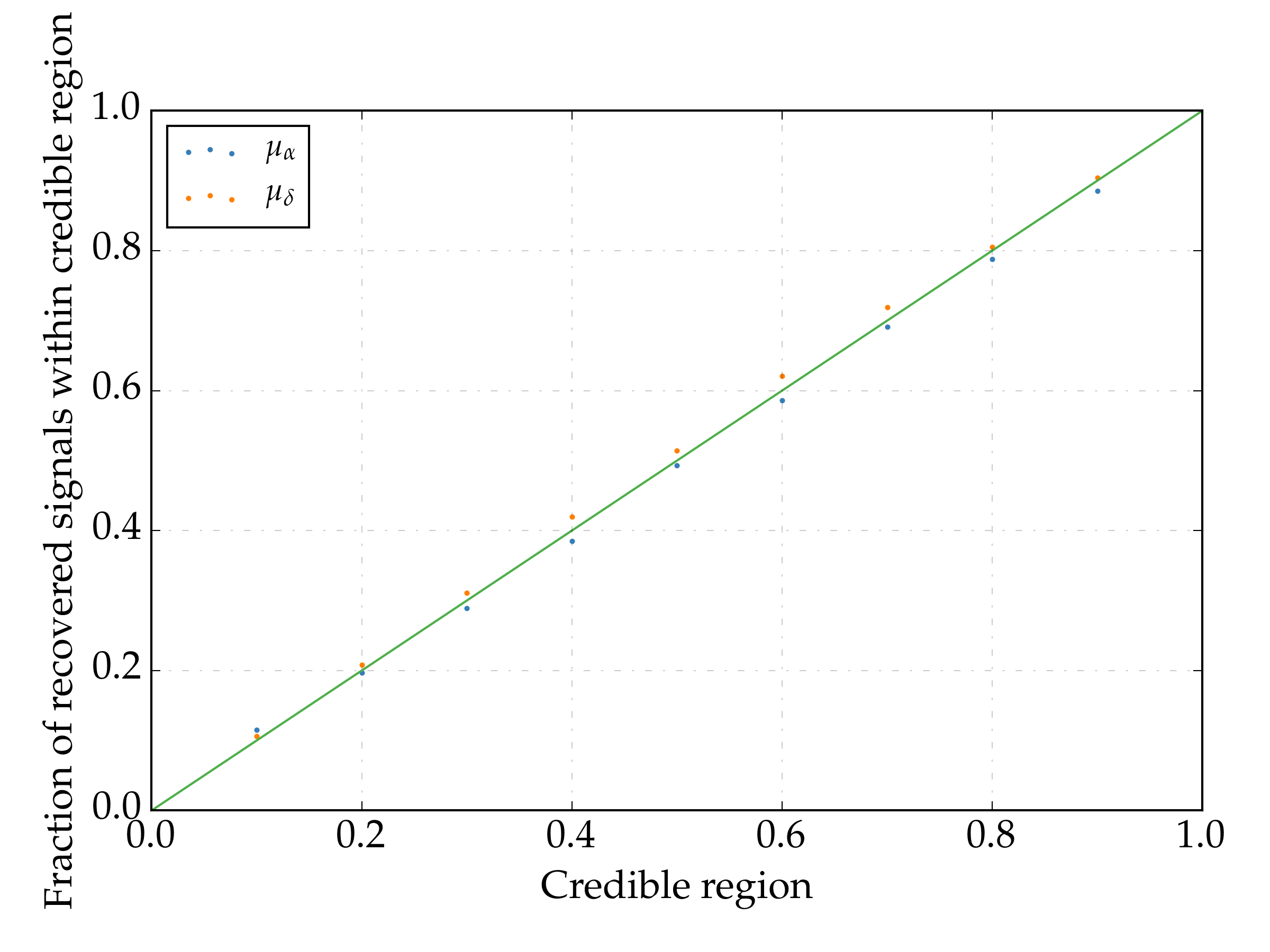}
\caption{Probability-probability plot showing the quantiles of the posterior distribution and the fraction of recovered signals whose true parameters are located inside the quantile.}
\label{fig:ppPlot}
\end{figure}

Firstly, we check that we recover the injections with the correct parameters. The pp-plot in figure \ref{fig:ppPlot} shows that we recover the proper motions of the injections within the expected credible regions (the same happens for all the other parameters describing the signal). This is calculated by taking the output chains of the MCMC procedure and calculating the credible region of each parameter. When the proper motion parameters are not searched, the obtained points do not follow the expected straight line (as expected from the biases found in subsection \ref{subsec:bias}), but when these parameters are included this figure shows that we recover the expected behaviour.

\begin{figure}
\includegraphics[width=1.0\columnwidth]{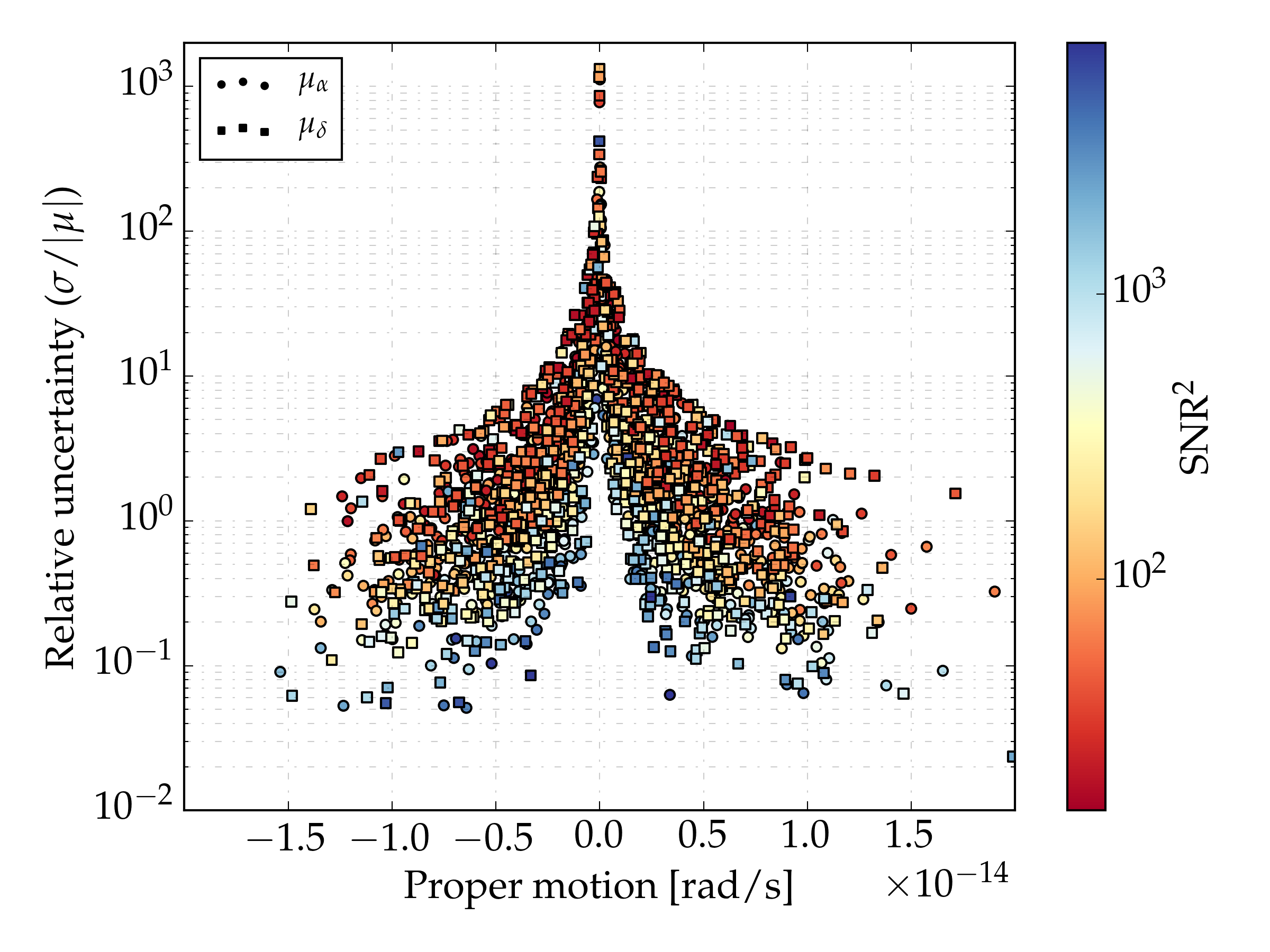}
\caption{Relative uncertainty (1-$\sigma$ divided by the true proper motion value) of the posterior distributions recovered with the MCMC follow-up scheme as a function of the proper motion value, for an all-sky coherent search of 2 years with 2 detectors. The color of each point (circles for the right ascension and squares for the declination) shows the squared signal-to-noise ratio of the signal.}
\label{fig:RelUnc}
\end{figure}

Secondly, we calculate the uncertainty of the recovered posterior distribution of the proper motion parameters. Figure \ref{fig:RelUnc} shows the relative uncertainty as a function of the proper motion and the squared signal-to-noise ratio of the signal, for a two years coherent search. As previously done, we inject signals at a range of SNRs, with isotropic orientations and sky distribution. It can be seen that, as expected, higher SNRs produce lower uncertainties, but other parameters not shown in this plot (such as the sky position or the frequency) also contribute to the vertical spread of similar proper motion values. The figure shows that for proper motions smaller than $10^{-15}$ rad/s a relative uncertainty smaller than 1 cannot be obtained (for a 2 years search with 2 detectors).

We have repeated the same injections for a 1 year search, and the theoretical increase of the uncertainty region following $T^{3/2}$ has been verified. Furthermore, we have also done a followup search where all the parameters except the proper motion ones were fixed to the true values, and the mean reduction of the relative uncertainty is around a factor of 2, meaning that we can get twice (on average) better uncertainty if the other parameters are known exactly and not searched. This could happen if for example a source is also found with an electromagnetic search.


From these results we remark that systematic calibration errors present in the current gravitational-wave data \citep{CalibrationUncertainty}, which are smaller than $5\%$, are narrower than our obtained relative uncertainties, at least for observation times less than a couple of years.



The accuracy on the proper motion parameters could be improved for some sky positions if ecliptical coordinates were used instead of equatorial coordinates, 
as for example discussed in \citep{PTA}. 
The metric components in ecliptical coordinates are obtained in the Appendix, and it can be seen that the accuracy that can be obtained at the same sky position is different for both coordinate systems.

\section{Conclusions}
\label{sec:conclusions}





In this paper we have studied the effects that proper motion of neutron stars produce on searches for continuous gravitational waves. All past searches have assumed the effect of proper motion to be negligible, but as we have seen this might be dangerous for coherent times longer than a year at frequencies higher than $\sim 1000$ Hz. Our results indicate that at these regions of parameter space, follow-up efforts should include these two extra parameters in the analysis, because otherwise a real signal could be missed. Since only a small number of outliers reach the final follow-up stage (where the number of segments is small) of searches for unknown neutron stars, this strategy should not highly increment the total computational cost of a search. This will be important for the upcoming observing runs that are planned to be longer than a year, or for when data from different observing runs is combined.

Besides the danger of missing a signal, we have also seen that even if this is not the case, the estimated parameters of the signal are biased. 
If these parameters are not included in a search, the reported uncertainties on the estimated parameters should be bigger in order to accomodate these systematic errors.

Finally, we have shown the relative uncertainty that can be achieved with a CW search of 2 years and 2 detectors, and how it depends on the SNR of the signal. Relative uncertainties smaller than 1 are only possible for proper motion values higher than $10^{-15}$ rad/s (for observing times less than a couple of years). A higher number of detectors would improve the relative uncertainty even more, due to the increase of the measured SNR.

In this paper we have not discussed the feasibility of measuring radial motion. As briefly mentioned in the signal model section, the effect of radial motion on the phase of the signal is smaller than the effect of transverse motion. Since we have seen that detecting transverse motion requires long observation times, measuring radial motion with CWs will be even more difficult and require observing times of many years.

We have assumed no timing noise or spin-wandering, which if present might bias the estimation of parameters in a similar way as the presence of glitches or proper motion. A detailed study of the size of these biases might be useful to uncover if these biases are bigger than the statistical uncertainties associated with our measurements. 

Another limitation of this study is that we have studied the bias introduced by the dismissal of proper motion in searches for isolated neutron stars. Biases introduced in searches for neutron stars in binary systems may be different, and the parameters describing the binary orbit (both Keplearian and post-Keplerian) might also be affected, as for example discussed in \citep{BinaryBiases}.

\section*{Acknowledgements}

The authors want to thank Karl Wette, Reinhard Prix, and David Keitel for many helpful discussions. We acknowledge the support of the Spanish Agencia Estatal de Investigaci{\'o}n and Ministerio de Ciencia, Innovaci{\'o}n y Universidades grants FPA2016-76821-P, FPA2017-90687-REDC, FPA2017-90566-REDC, FPA2015-69815-REDT, FPA2015-68783-REDT, the Vicepresidencia i Conselleria d'Innovaci{\'o}, Recerca i Turisme del Govern de les Illes Balears and the Fons Social Europeu 2014-2020 de les Illes Balears, the European Union FEDER funds, and the EU COST actions CA16104, CA16214 and CA17137. The authors are grateful for computational resources provided by the LIGO Laboratory and supported by National Science Foundation Grants PHY-0757058 and PHY-0823459. This article has LIGO document number P2000238.




\bibliographystyle{mnras}
\bibliography{main_MNRAS} 

\appendix

\section{Explicit derivation of metric components}

In this appendix we explicitly derive the $g_{\mu_{\alpha} \mu_{\alpha}}$ metric component shown in section \ref{sec:mismatch} as an example. The other metric components can be obtained in a similar way as the one showed below.

The phase in the detector frame is given by (up to first-order in frequency derivatives, and assuming that $f_k \approx f'_k$):
\begin{align}
    \phi(t) &= \phi_0 + 2\pi f_0 \left[ (t-t_r) + \frac{\vec{r}(t)\cdot\hat{n}(t)}{c}\right] \nonumber \\ & + \pi f_1 \left[(t-t_r)^2 + \left( \frac{\vec{r}(t)\cdot\hat{n}(t)}{c}\right)^2 + 2(t-t_r)\frac{\vec{r}(t)\cdot\hat{n}(t)}{c}\right],
\end{align}

The first step consists on calculating the phase derivative with respect to the parameter:
\begin{align}
    \frac{\partial \phi}{\partial \mu_{\alpha}} &= \frac{2\pi f_0}{c} \vec{r}(t)\cdot \frac{d\hat{n}(t)}{d\mu_{\alpha}} \nonumber \\ &+ \pi f_1 \left[2 \left(\frac{\vec{r}(t)}{c}\right)^2\hat{n}(t)\cdot\frac{d\hat{n}(t)}{d\mu_{\alpha}} + \frac{2(t-t_r)}{c}\vec{r}(t)\cdot \frac{d\hat{n}(t)}{d\mu_{\alpha}}\right],
\label{eq:phder}
\end{align}
where $\frac{\partial \hat{n}(t)}{\partial \mu_{\alpha}} = (t-t_r)[-\sin{\alpha_0}\cos{\delta_0}, 
 \cos{\alpha_0}\cos{\delta_0}, 0]$.

The relative importance of the different terms can be estimated with the following order of magnitude calculation (separating $\vec{r}(t)$ in its two contributions):
\begin{align}
   \frac{\partial \phi}{\partial \mu_{\alpha}} \sim &\frac{f_0 T R_{ES}}{c} + \frac{f_0 T R_{E}}{c} + \frac{f_1 T R_{ES}^2}{c^2} + \frac{f_1 T^2 R_{ES}}{c} \nonumber \\
   & + \frac{f_1 T R_{E}^2}{c^2} + \frac{f_1 T R_{ES} R_{E}}{c^2} + \frac{f_1 T^2 R_{E}}{c}.
\end{align}
For realistic values of $f_0$ and $f_1$, and for integration times shorter than several years, the terms with $f_0$ are always much bigger that the terms with $f_1$ and higher-order frequency derivatives. 
Furthermore, this estimation shows that we can approximate the derivative by just keeping the terms dependent on $R_{ES}$, thus neglecting the rotation of the Earth.
 
The last step consists on calculating the time integrals:
\begin{align}
    \langle \frac{\partial \phi}{\partial \mu_{\alpha}} \frac{\partial \phi}{\partial \mu_{\alpha}} \rangle = \frac{1}{T} \int_{t_0}^{t_0+T} \left(\frac{2\pi f_0}{c} \vec{r}_O(t)\cdot \frac{\partial \hat{n}'(t)}{\partial \mu_{\alpha}}\right)^2 dt  \nonumber \\
    = \frac{4\pi^2 f_0^2 R^2_{ES}}{T c^2} \int_{t_0}^{t_0+T} (t-t_r)^2 ( -\sin{\alpha_0}\cos{\delta_0}\cos{[\phi_O + \Omega_O (t-t_r)]} \nonumber \\
    +  \cos{\alpha_0}\cos{\delta_0}\cos{\epsilon}\sin{[\phi_O + \Omega_O (t-t_r)]} )^2 dt \nonumber \\
    \approx \frac{4\pi^2 f_0^2 R^2_{ES}}{T c^2}\left[ \right. (\sin^2{\alpha_0}\cos^2{\delta_0}+\cos^2{\alpha_0}\cos^2{\delta_0}\cos^2{\epsilon}) \nonumber \\ \left(  \frac{(T+t_0)^3}{6} \right. - \frac{t_r(T+t_0)^2}{2} + \frac{t_r^2(T+t_0)}{2} - \frac{t_r^2 t_0}{2} + \frac{t_r t_0^2}{2} - \frac{t_0^3}{6} \nonumber \\
    + \mathcal{O} \left. \left(\frac{T^2}{\Omega_O}\right) + \mathcal{O}\left(\frac{T}{\Omega_O^2}\right) + \mathcal{O}\left(\frac{1}{\Omega_O^3}\right)\right) \nonumber \\ - (\sin{\alpha_0}\cos{\delta_0}\cos{\alpha_0}\cos{\delta_0}\cos{\epsilon})\left.\left(\mathcal{O}\left(\frac{T^2}{\Omega_O}\right) + \mathcal{O}\left(\frac{T}{\Omega_O^2}\right) + \mathcal{O}\left(\frac{1}{\Omega_O^3} \right) \right)     \right] \\
    \langle \frac{\partial \phi}{\partial \mu_{\alpha}} \rangle \langle \frac{\partial \phi}{\partial \mu_{\alpha}} \rangle = \frac{1}{T^2} \left( \int_{t_0}^{t_0+T} \frac{2\pi f_0}{c} \vec{r}_O(t)\cdot \frac{\partial \hat{n}'(t)}{\partial \mu_{\alpha}} dt \right)^2 \propto \mathcal{O}(\frac{T}{\Omega_O})
\end{align}
It can be seen that for integration times longer than a year, terms such as $T^2/\Omega_O$ are smaller than $T^3$.

Now, we fix the reference time to two different values. For $t_r= t_0 + T/2$:
\begin{align}
  &\frac{(T+t_0)^3}{6} - \frac{t_r(T+t_0)^2}{2} + \frac{t_r^2(T+t_0)}{2} - \frac{t_r^2 t_0}{2} + \frac{t_r t_0^2}{2} - \frac{t_0^3}{6} = \frac{T^3}{24} \nonumber \\
  &g_{\mu_{\alpha} \mu_{\alpha}} = \frac{4\pi^2 f_0^2 R^2_{ES} T^2}{24 c^2} (\sin^2{\alpha_0}\cos^2{\delta_0}+\cos^2{\alpha_0}\cos^2{\delta_0}\cos^2{\epsilon}) + \mathcal{O}(T/\Omega_O)
\end{align}
while for $t_r=t_0$ or $t_r=t_0 + T$:
\begin{align}
  &\frac{(T+t_0)^3}{6} - \frac{t_r(T+t_0)^2}{2} + \frac{t_r^2(T+t_0)}{2} - \frac{t_r^2 t_0}{2} + \frac{t_r t_0^2}{2} - \frac{t_0^3}{6} = \frac{T^3}{6} \nonumber \\
  &g_{\mu_{\alpha} \mu_{\alpha}} = \frac{4\pi^2 f_0^2 R^2_{ES} T^2}{6c^2} (\sin^2{\alpha_0}\cos^2{\delta_0}+\cos^2{\alpha_0}\cos^2{\delta_0}\cos^2{\epsilon}) + \mathcal{O}(T/\Omega_O)
\end{align}
Reference times selected between the initial and mid-time will produce metric components with values between these two extremes. 
With this approximation and equation \eqref{eq:metricdef} we obtain the metric elements shown in section \ref{sec:mismatch}. 


In ecliptical coordinates where $l$ is the longitude and $b$ is the latitude, we define the source and Earth positions:
\begin{align}
\hat{n}'(t) &= \hat{n}(t_r) + \hat{\dot{n}}(t_r)(t-t_r) \nonumber \\
 &= [\cos{l_0}\cos{b_0},\sin{l_0}\cos{b_0},\sin{b_0}] \nonumber \\ 
 & + (t-t_r)[-\mu_{l}\sin{l_0}\cos{b_0} - \mu_{b}\cos{l_0}\sin{l_0}, \nonumber \\
 & \mu_{l}\cos{l_0}\cos{b_0} - \mu_{b}\sin{l_0}\sin{b_0}, \mu_{b}\cos{b_0}] \\
  \vec{r}_O (t) &= R_{ES}[\cos{\left(\phi_O + \Omega_O (t-t_r)\right)}, \nonumber \\ & \sin{\left(\phi_O + \Omega_O (t-t_r)\right)}, 0] .
\end{align}
When the time integrals are done in these coordinates, the results are (for $t_r=t_0 + T/2$):
\begin{align}
    g_{\mu_{l}\mu_{l}} &= \frac{4\pi^2 R_{ES}^2 f_0^2 T^2}{24c^2} \cos^2{b} + \mathcal{O}(T/\Omega_O), \nonumber \\
    g_{\mu_{b}\mu_{b}} &= \frac{4\pi^2 R_{ES}^2 f_0^2 T^2}{24c^2} \sin^2{b} + \mathcal{O}(T/\Omega_O), \nonumber \\
    g_{\mu_{l}\mu_{b}} 
    &\propto \mathcal{O}(T/\Omega_O), \nonumber \\
\end{align}
It can be seen that in these coordinates the covariant component is reduced as compared to the equatorial coordinates case.


\bsp	
\label{lastpage}
\end{document}